\documentclass[sigconf]{acmart}

\usepackage{url}  
\usepackage{graphicx}  
\usepackage{multirow}

\usepackage{amsmath}
\usepackage[skip=10pt]{caption}

\usepackage{balance}

\begin{document}
\title{Scalable Knowledge Graph Construction from Twitter }
\author{Omar Alonso, Vasileios Kandylas, Serge-Eric Tremblay,\\
Microsoft\\
}

\begin{abstract}
We describe a knowledge graph derived from Twitter data with the goal of discovering relationships between people, links, and topics. The goal is to filter out noise from Twitter and surface an inside-out view that relies on high quality content. The generated graph  contains many relationships where the user can query and traverse the structure from different angles allowing the development of new applications.
\end{abstract}

\maketitle

\section{Introduction}

Social networks continue to be a top destination for information consumption on the Internet. The ever-expanding social graph based on friends (Facebook) or followers (Twitter) enables the implementation of traditional features like recommendations (e.g., links, friends, accounts, etc.) and trending topics that rely on human input and other behavioral data. That said, given the enormous amount of human sensing in the world at any given moment in any of those products, there seems to be a lot of untapped potential that goes beyond simple applications on top of atomic level content like a post or tweet. 

We believe that by mining and extracting specific content from inside a social network, a knowledge graph can be automatically derived and later be used as underlying structure for a number of applications. 
Knowledge bases (KB) contain rich semantic information about entities and relationships, and have become a very powerful asset in web search, with Google's Knowledge Graph and Microsoft's Satori  being two prominent examples. KBs usually extract information from web pages or Wikipedia and they are built on existing and specific relation types. Recall tends to be a problem with KBs as available content covers specific types and facts.

Twitter, in contrast, produces real-time information which can be extremely valuable for detecting new relationships that would take some time to have an entry on Wikipedia or other web property. At the same time, Twitter is noisy and given the textual limitation, traditional extraction approaches tend not to work well. 
 In our research, we differ  from traditional KB construction and focus on the efficient and scalable generation of a graph with semantic annotations that suits better the characteristics of social network data.
In this paper, we  present the design and implementation of an unsupervised approach that uses the Twitter firehose as the only data input
and produces as output a Social Knowledge Graph (SKG). More precisely, 
we describe techniques for: 1) efficient mining of large scale social data for extraction of links, trusted users, and topics, 2)
computation of relationships, contextual vectors, signatures, and entity stamping, and 3)
population of the SKG schema. 

\section{Social Knowledge Graph}

Compared to previous work on KB generation and information extraction from Twitter, we take a bottom-up approach with an emphasis on identifying good quality elements first. Selecting good content, users and links in an efficient manner enables the creation of connections for high quality elements. 
  The utility of this SKG is to retrieve, extract, and present social information as a unit that can be beneficial to many applications. Table \ref{tab:skg-pivots} summarizes the main components in SKG. 

\begin{table}[h!]
  \centering
  \begin{tabular}{| p{1.6cm} | p{6cm} |} 
    \hline
\textbf{Node types} & \textbf{Description}   \\ \hline     
Users	&We rely on a set of Twitter users called  trusted users which are discovered by 2-way communications initiated by  verified  users.  \\ \hline
Links		&Popular links that are shared by those trusted users. \\ \hline
Topics		&Extraction of topics such as entities, hashtags, and n-grams from tweets. \\ \hline
Posts	&Posts are tweets used  as supporting evidence (or provenance) for each node and connection in the graph.\\ \hline
Time &The graph is archived at regular intervals with snapshots  which can be used to produce a timeline view of key topics. \\ \hline
  \end{tabular}
  \caption{SKG high-level overview.} 
  \label{tab:skg-pivots}
\end{table}

We briefly describe a number of techniques that we use as building blocks on the graph. 
A \emph{social signature} is a high level representation of a web page from a social media perspective, that is, a list of n-grams associated with the link.  For computing a social signature, we use the text of the tweets that share the link and aggregate all the social anchor text around that link from different users. The social anchor text consists of the text in the tweet, excluding the Twitter user profile handles. A score is calculated for each candidate social signature using a learned model that computes a weighted combination of the following features: term frequency (tf), document frequency (df), term frequency-inverse document frequency (tf-idf) of the n-gram calculated as 
${\text{tf-idf}} = {1+\log(tf)} * \log\left(\frac{{n}}{{df}}\right)$,  and  local/global affinity of the n-gram.

A \emph{contextual vector} represents a ranked list of n-grams for a set of tweets related to a hashtag or entity.
For producing the list of n-grams, we first aggregate all the tweets related to a particular hashtag over the time period of consideration. 
The computation of the contextual vectors is similar to that of the social signatures. The main difference is that contextual vectors are computed for hashtags or entities instead of links. Small differences in the computation account for the different characteristics of hashtags and entities vs. links and weight scoring.

A user is considered \emph{trusted} if they had 2-way communication (@mention) with a Twitter verified or already trusted user, initiated by the verified or trusted user~\cite{HentschelACK14}. This computation is repeated, expanding the set of trusted users by another ring in each iteration. In practice, no more users are added after the $10^{th}$ repetition.

\section{Schema and Knowledge Graph Population}

The SKG schema captures the main components of the social network and the connections between them. It focuses on ``first order'' connections, i.e. those immediately discoverable. More complicated connections, which can be derived by more advanced analysis of the data, are left to be computed by other applications based on the SKG data and as required by the application.

The design philosophy is that the data are computed on a continuous basis (e.g., hourly) and capture the main information from the social network, but they do not try to do everything. The SKG data are the building blocks and specific applications are built on top of them. The applications read the SKG data, compute more complicated or advanced relationships (e.g., topic timelines), combine any data from other sources and derive application-specific data for their own use.

\begin{figure}[h]
\begin{center}
\includegraphics[scale=0.15]{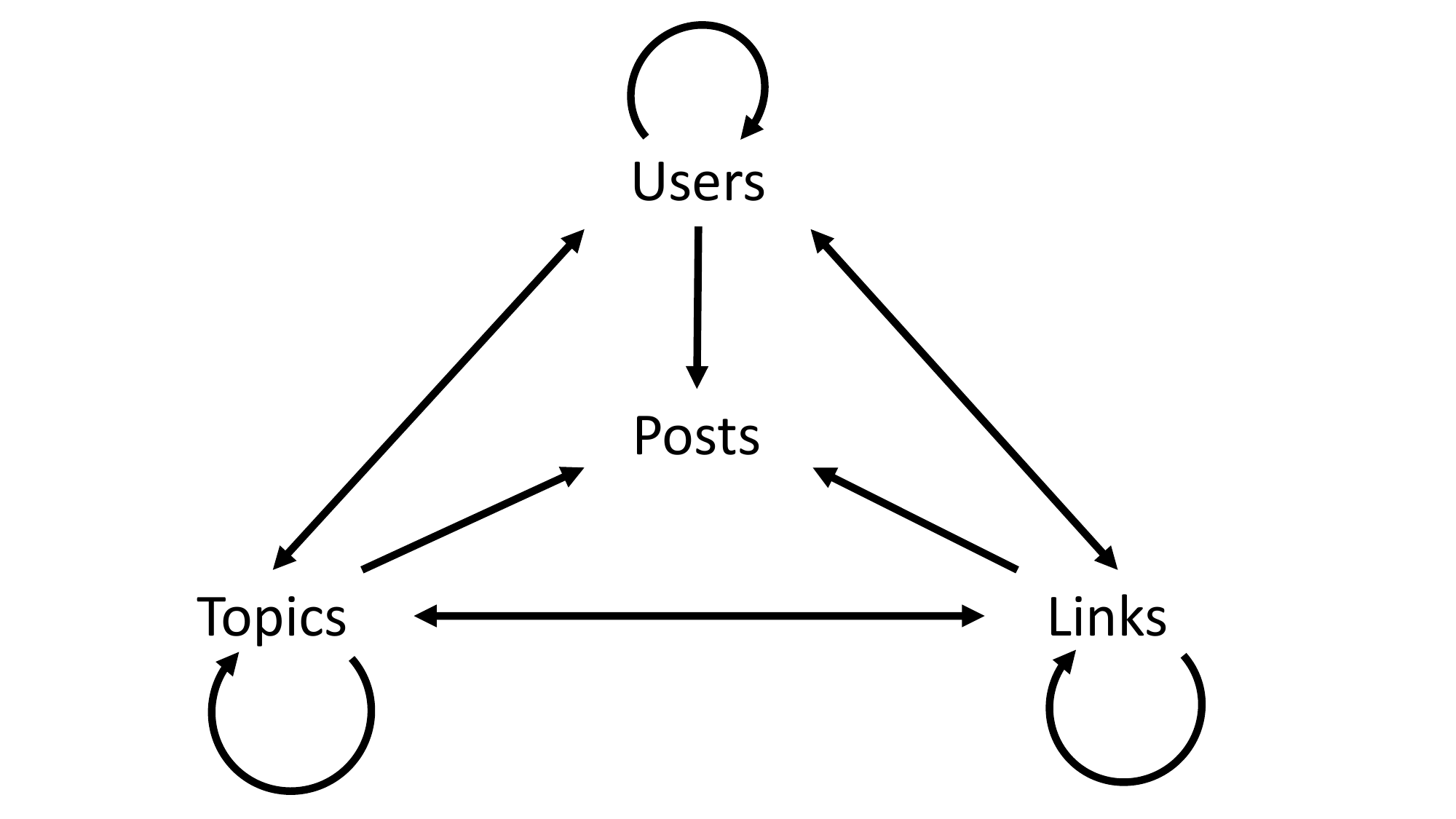} 
\caption{SKG schema.}
  \label{fig:schema}
\end{center}
\end{figure}

The SKG schema consists of 4 component tables, which contain information about the nodes of the graph, and 9 connection tables, which link the nodes of the graph and contain any extra information needed to select the desired types of connections (Figure~\ref{fig:schema}). 
The component tables are:
\begin{enumerate}
\setlength{\parskip}{0pt}
\item \texttt{Users}: Contains information about the user (name, profile picture, authority score, number of followers, etc.).
\item  \texttt{Links}: Contains information about the links mentioned/shared in the posts (URL, title, description, spam score, etc.).
\item  \texttt{Topics}: Contains information about the entities, hashtags, cashtags (e.g. \$MSFT), or n-grams mentioned in the posts (topic type, frequency, context vector etc.).
\item  \texttt{Posts}: Contains a small number of top posts to be used as supporting evidence for the nodes and edges of the graph and their information (timestamp, quality score, retweets, favorited count, etc.).
\end{enumerate}

The connection tables contain information about the connection between two elements. They are:
\begin{enumerate}
\setlength{\parskip}{0pt}
\item  \texttt{Users-Links}: user-link relation (authored, coreferenced), score, supporting posts.
\item      \texttt{Users-Topics}: topic type, user-topic relation (authored, coreferenced, fanOrAuthority), score, supporting posts.
\item       \texttt{Users-Users}: user-user relation (mentioned, retweeted, replied, coreferenced), score, supporting posts.
\item     \texttt{Links-Topics}: topic type, score, supporting posts.
\item     \texttt{Topics-Topics}: topic type, score, supporting posts.
\item     \texttt{Links-Links}: score, supporting posts.
\item     \texttt{Users-Posts}: timestamp, relation (authored, mentioned), score.
\item     \texttt{Links-Posts}: timestamp, score.
\item     \texttt{Topics-Posts}: topic type, timestamp, score.
\end{enumerate}

The topics are always annotated by their type (hashtag, n-gram, entity, cashtag), so that  it is easy to select, for instance only the hashtags associated with a user.

Bidirectional connections are indexed in both directions, both by the first and by the second element of the connection (e.g., the table \texttt{Links-Topics} is indexed both by link and by topic). This allows fast lookups and connecting from either direction. Other than that, the two directions have identical information. Connections of any element with topics, however, have differences between the two directions. When the topics are the second element of the connection, they are clustered, but when they are the first, they are not. The topic clustering allows searching for connections, such as links associated with the topic, using only parts of a topic (e.g., ``Obama'' for the topic ``Barack Obama''), but when the topics are returned as results, only the full topic is returned (e.g., the n-gram ``Barack Obama'') and not its parts (not ``Obama'').

Most connection tables contain a description of the relation represented by the connection. When a user is involved, the relation could be: authored (the user authored the post that contains the topic, link, etc.), coreferenced (the user was mentioned in the same post together with the other topic, link, etc.) or mentioned (the user was mentioned in the post). For \texttt{Users-Topics}, the user could also be a fan or authority on the topic, discovered with an expertise detection algorithm. For \texttt{Users-Users}, the relation captures if one user mentioned the other in the post, replied to the other user, retweeted the other user, or if both users were coreferenced in the same post.

All tables contain one main score. Other scores are also available depending on the table. For the component tables, the main score is used for static ranking users, topics, links, and posts. This score is computed from a combination of other scores that may be available, such as the frequency of occurrence, spam score, authority score etc. For the connection tables, the main score represents the strength of the connection and is based on a normalized frequency of occurrence. The normalization is enforced by allowing every account to have a single ``vote'' for the connection. So, repeated posts by the same account, mentioning for instance  a link and a topic, only contribute one vote to the connection of that link and topic. This helps reduce the effect of spammer and advertising accounts who, by repeatedly tweeting the same things, would otherwise over-inflate the importance of the connection. 

\subsection{Data selection and computations}

The SKG schema does not store all the social network data, but only a subset. This section describes how this subset is selected.

First, all the posts are filtered by using appropriate spam, authority, profanity and adult score thresholds, both for the text of the post and any links in the post. Then, the individual components of the graph are selected.
The users are  filtered by selecting users who are non-protected, verified, or trusted  and satisfy appropriate user authority, adult and followers count thresholds.
 Trusted users increase user coverage (compared to verified users), while maintaining high quality users (non-spammers).
The links selected for the graph are:
\begin{itemize}
\setlength{\parskip}{0pt}
\item Top popular links (most often shared).
\item Top viral links using the average pairwise distance between nodes in the
diffusion tree on $\nu(T) = \frac{1}{n(n-1)} \sum_{i = 1}^n \sum_{\substack{j=1\\ j\ne i}}^n d_{ij},  $ where $d_{ij}$ indicates the length of the shortest path between nodes
$i$ and $j$, as proposed by Goel et al.~\cite{GoelAHW16}.
\item Top trending links (detected by a proprietary trending detection algorithm)
\end{itemize}

The hashtags, cashtags, entities and n-grams (from the post text after removal of links, mentions and stop words) are extracted. The selected topics are:

\begin{itemize}
\setlength{\parskip}{0pt}
\item Top popular and trending hashtags, cashtags, entities, n-grams.
\item Link social signature n-grams (for each link, compute the social signatures, then add the social signature n-grams).
\item Hashtag and entity contextual vector n-grams.
\end{itemize}

After the above selections, the connection tables are generated. These select all the connections where both sides (e.g. user and link) appear in the above selected subsets. \texttt{Users-Users} connections are selected based on the number of times a user mentions, replies, retweets another user, or both users are co-referenced in the same post. 
\texttt{Topics-Topics} and \texttt{Links-Links} connections are selected based on the number of times the two topics or two links co-occur in a post.
Finally, the posts are selected. Posts are considered a secondary component of the SKG schema. They are added to provide examples for any selected user, link or topic. For example, for a topic in the graph, the selected posts are examples of posts that mention the topic. For any user, link or topic selected for the graph, a small number of the top associated posts (according to post score) are selected as examples and all connections to the user, link or topic are added. Associated posts are those that were written by the user, or mention/$@$/RT the user, or contain the link or topic.

\begin{figure*}[ht!]
\begin{center}
\includegraphics[scale=0.29]{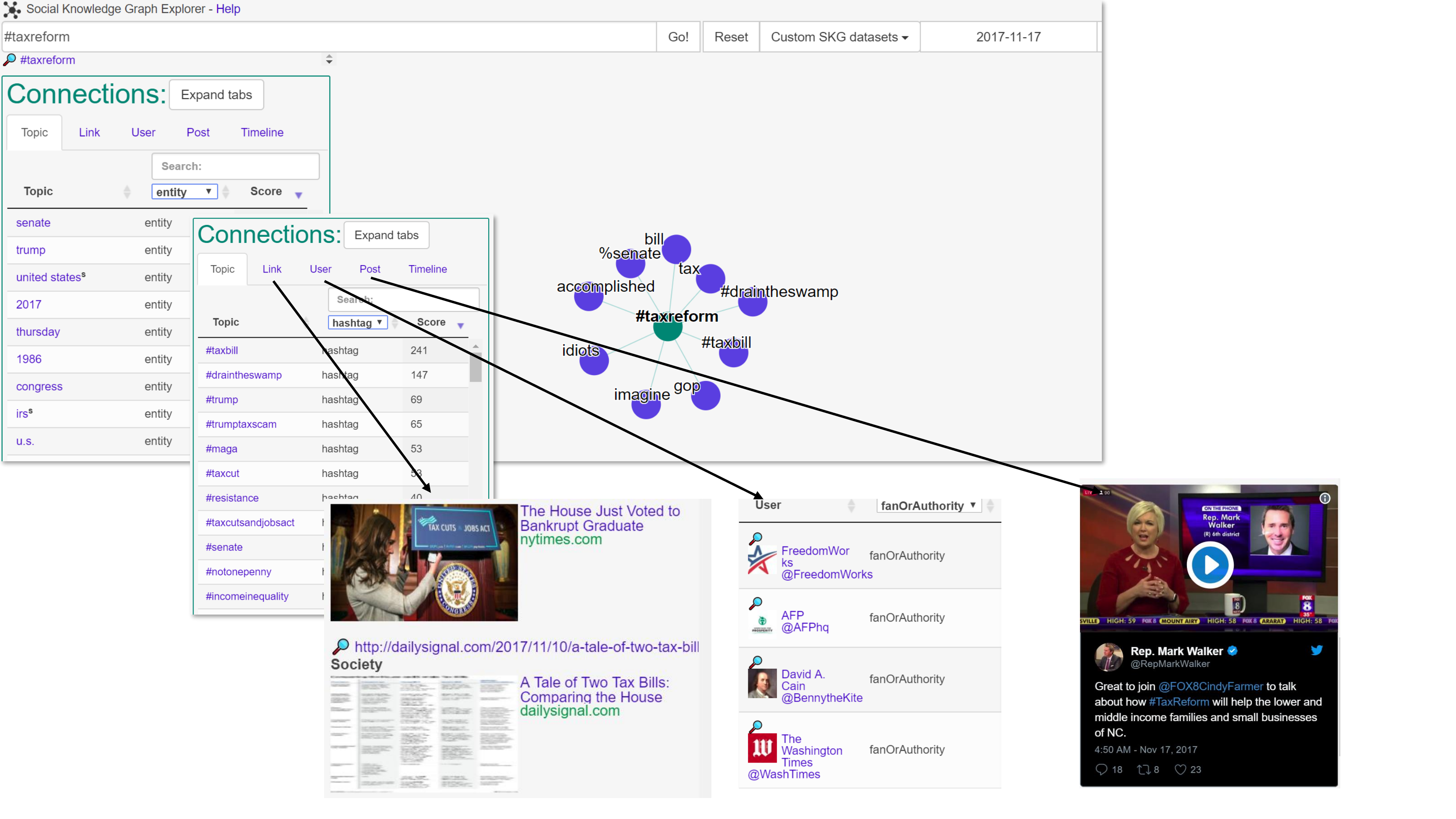} 
\caption{The SKG explorer application. We search for a  topic (tax reform) on a specific day (November 11, 2017) and filter the connections by entities (e.g., Senate, Trump, USA, etc.) and hashtags (e.g., \#taxbill, \#taxcuts, etc.). Clicking on the ``Link'' tab shows a few examples of links shared on Twitter.  Same goes for the tabs ``User'' and ``Post''. The visualization in the middle pane allows us to explore the graph visually by clicking on the nodes.}
  \label{fig:skg-explorer}
\end{center}
\end{figure*}

\subsection{Temporal resolution}

The previously mentioned computations are executed on an hourly basis, with each execution processing the messages posted in the most recently available one-hour interval. Each produced graph is thus a snapshot of the social activity during that hour. To analyze the activity over a longer time interval (e.g., a day or a week), we could apply the same approach to messages from the longer interval, but this is inefficient, both because it would require more processing time and because all these messages have already been processed. So instead, we developed a process to aggregate and summarize the graphs of multiple, shorter time intervals into a single graph corresponding to a longer time interval. In our case, at the end of each day, we aggregate the 24 hourly graphs into a single daily summary graph. This is repeated hierarchically, so at the end of each month we aggregate all the daily graphs of the month, and at the end of the year we aggregate all monthly graphs to produce a yearly summary graph. In order to do the aggregation of the graph tables, we take the union of users, links, topics, posts and connections. For the matching elements (e.g., the same link, or the same topic-user connection), the various frequency and vote scores from all hours are being added. Where it makes sense, we don't use summation but a more appropriate operation. For the virality scores we use the maximum value, for dates and timestamps we generally use the earliest occurrence, and for scores that become more accurate with time (such as the spam and authority scores which are more accurate with more available data), we use the most recent value. For supporting evidence and to avoid continuous accumulation of posts, we keep a fixed amount of posts per component or connection, by selecting the highest scoring ones within the longer time interval. 

The above process generates an aggregate graph that contains every component and connection of the individual graphs. However, in practice, some tail elements are not worth keeping. For instance, a topic or link might have high tweet frequency during some hour, but die down quickly in later hours. Such elements, resembling sparks that burn and fizzle quickly, when aggregated will have very low daily frequencies. And as we compute monthly and yearly graphs, these tail elements will just accumulate and pollute the graph, without serving any practical purpose. Therefore, we filter out such elements whose total votes and/or frequencies are below appropriately selected thresholds. The thresholds are chosen based on a combination of an empirically tuned value depending on the type of the element (e.g., there are different values for hashtags, ngrams, user-link mentions, link-hashtag cooccurrences etc.) and a time interval multiplier which depends on the the total length of the time intervals being aggregated and accounts for the longer duration and accordingly the higher expected number of occurrences. The produced graph can then be thought of as a summary of the longer time interval that contains all the salient elements but not other spurious events of  smaller importance.

\section{Results and Evaluation}

The complete back-end data pipeline and algorithms have been implemented in a MapReduce-like framework and run on a distributed cluster  continuously. We use our own
in-house NER implementation trained on tweets for detecting people, places, and organizations. Figure~\ref{fig:skg-explorer} shows a visual explorer for SKG for the tax reform topic the day after the bill was passed. The user can dissect the content in the Connections tabs and sort data by filters, use the visualization to discover relationships, or enter a query for a topic or entity on the search box.

We generated SKG using 2 years worth of Twitter data. The underlying data processing pipeline  processes 120M tweets daily filtered by spam/adult/etc. We keep only those tweeted by our 15M trusted users which leaves around 22M tweets. We select 65K links, 600K topics, and 65K hashtags according to tunable thresholds which may change. We then compute many connections, for example, 
3.5M user-user interactions and 100K topic-topic relations. Finally, we select 5 tweets per user,  per topic and  per link for a total of 6M tweets. This amounts to around 11 Gigabytes daily.

Due to confidentiality, however, we do not disclose user engagement  or other behavioral data. Instead, we present an offline relevance evaluation.
We conducted an offline evaluation of 100 links selected at random with an internal tool for collecting labels using in-house editors. Each element was judged by 5 editors and we took majority vote as the final label. If the majority vote says that the entry is negative, then there is a defect. Figure~\ref{template} shows the relevance evaluation template and Table~\ref{tab:offline} contains the defect rate for the three evaluation questions.

\begin{figure}[h!]
\begin{center}
{\small 
\fbox{\begin{minipage}{0.92\linewidth}
\textit{Please help us evaluate the relevance of an article (or link) and associated information (e.g., image, hashtag, etc.). You will be given a recent article that has been shared by a user in a social network like Twitter along with its related hashtag. Your task is to assess the quality of each item in the 3 questions below.}

\texttt{}

[Show \$Link title\$; \$Image\$; \$Hashtag\$]

\texttt{} 

Q1. Do you think the article should be of interest to the majority of users?

[] Yes [] Somewhat [] No

Q2. Is the image thumbnail relevant to the article?

[] Yes [] Somewhat [] No

Q3. Is the related hashtag relevant to the article?

[] Yes [] Somewhat [] No

\end{minipage}}
}
\end{center}
\caption{Offline task relevance evaluation template.}
\label{template}
\end{figure}

\begin{table}[h!]
  \centering
  \begin{tabular}{| c | c | }
    \hline
\textbf{Question} & \textbf{Defect rate}   \\ \hline     
Q1	&1\%	\\ \hline
Q2	&3\%	 \\ \hline
Q3	&6\%	 \\ \hline
  \end{tabular}
  \caption{Defect rates for the offline evaluation.} 
  \label{tab:offline}
\end{table}

\section{Related Work}

There is little published work on extracting knowledge or creating knowledge graphs using the entire Twitter firehose as input.
One of the first  information extraction systems from Twitter is TwiCal, that categorizes events from tweets \cite{RitterMEC12}. Knowledge extraction using frame semantics from Twitter on a curated data set is described in 
\cite{SogaardPA15}. Weikum et al.~\cite{WeikumHS16} identify social data as good source for knowledge extraction.

\section{Conclusion}

We described the design and implementation of SKG, a social knowledge graph derived automatically from Twitter. Our techniques scale to the Twitter firehose and have been deployed in a production system. In our work, the focus is on data and 
information quality as preconditions for identifying relationships. SKG allows users to query the graph by topic, user, or link, and a temporal dimension is also added for anchoring relationships in time.
The same techniques can be used to produce on-demand SKGs. That is, specialized Topic/User/Links graphs around different groups of users such as professional soccer players, Olympians, top CEOs, etc. This technology enables us to find the most relevant content from the perspective of a specific set of users.

\bibliographystyle{ACM-Reference-Format}
\bibliography{skg-refs}

\end{document}